\documentclass[preprint,12pt]{elsarticle}
\usepackage{graphicx} 

\usepackage{amssymb}
\usepackage{amsmath}
\usepackage{xcolor}
\usepackage{url} 
\usepackage[colorlinks=true]{hyperref}

\usepackage[utf8]{inputenc}
\usepackage[T1]{fontenc}      

\makeatletter
\def\@fnsymbol#1{\ensuremath{\ifcase#1\or *\or \dagger\or \ddagger\or
  \mathsection\or \mathparagraph\or \|\or **\or \dagger\dagger\or \ddagger\ddagger\fi}}
\makeatother

\begin{document}

\begin{frontmatter}

\def\@fnsymbol#1{\ensuremath{\ifcase#1\or *\or \dagger\or \ddagger\fi}}
 
\title{Rapidly Spinning Massive Pulsars as an Indicator of Quark Deconfinement}
 
 \author{Christoph Gärtlein\fnref{label1,label2,label3,1}}

 \fntext[1]{e-mail: christoph.gartlein@tecnico.ulisboa.pt}

 \affiliation[label1]{organization={Centro de Astrofísica e Gravita\c c\~ao  - CENTRA, Departamento de F\'{\i}sica, Instituto Superior T\'ecnico - IST, Universidade de Lisboa - UL},
             addressline={Av. Rovisco Pais 1},
             city={Lisboa},
             postcode={1049-001},
             country={Portugal}}

\affiliation[label2]{organization={CFisUC, Department of Physics, University of Coimbra},
city={Coimbra},
postcode={3004-516},
country={Portugal}}

\affiliation[label3]{organization={Institute of Theoretical Physics, University of Wroclaw},
city={Wroclaw},
postcode={50-204},
country={Poland}}

\author{Violetta Sagun\fnref{label4,2}}

\fntext[2]{e-mail: v.sagun@soton.ac.uk}
 
\affiliation[label4]{organization={Mathematical Sciences and STAG Research Centre, University of Southampton},
city={Southampton},
postcode={SO17 1BJ},
country={United Kingdom}}
  
\author{Oleksii Ivanytskyi 
\fnref{label5,3}}
\fntext[3]{e-mail: oleksii.ivanytskyi@uwr.edu.pl}

\affiliation[label5]{organization={Incubator of Scientific Excellence---Centre for Simulations of Superdense Fluids, University of Wroclaw},
    city={Wroclaw},
             postcode={50-204},
             country={Poland}}
 
\author{David Blaschke\fnref{label3,label6,label7,4}}

\fntext[4]{e-mail: david.blaschke@uwr.edu.pl}

\affiliation[label6]{organization={Helmholtz-Zentrum Dresden-Rossendorf (HZDR)},
             addressline={Bautzner Landstrasse 400},
    city={Dresden},
             postcode={01328},
             country={Germany}}

\affiliation[label7]{organization={Center for Advanced Systems Understanding (CASUS)},
             addressline={Untermarkt 20},
    city={G\"orlitz},
             postcode={02826},
             country={Germany}}

\author{Ilídio Lopes\fnref{label1,5}}
\fntext[5]{e-mail: ilidio.lopes@tecnico.ulisboa.pt}


\begin{abstract}
We study rotating hybrid stars, with particular emphasis on the effect of spin on the deconfinement phase transition and star properties. Our analysis is based on a hybrid equation of state with a phase transition from hadronic matter containing hyperons to color-superconducting quark matter, where the quark phase is modeled within a relativistic density functional approach. By varying the strength of the vector repulsion and diquark pairing couplings in the microscopic quark Lagrangian, we construct a set of hybrid star sequences with different quark-matter onset densities. This framework ensures consistency with astrophysical and gravitational wave constraints on mass, radius, and tidal deformability.

We show how increasing the rotational frequency affects the maximum gravitational mass, oblateness, central energy density of compact stars, as well as the onset of quasiradial oscillations and nonaxisymmetric instabilities. Our results indicate that, for the most favorable parameter sets with strong vector coupling, hybrid stars containing a color-superconducting quark-matter core can account for the fastest-spinning and most massive known neutron star, PSR J0952–0607. In contrast, this pulsar cannot be explained by the purely hadronic equation of state considered in this study.

These findings are highly relevant for future observations with the Square Kilometre Array (SKA), which will significantly expand the known pulsar population and enable precise measurements of masses and spin frequencies.

\end{abstract}


\end{frontmatter}



\section{Introduction}
\label{int}

\noindent Nature's most precise clocks, as millisecond pulsars (MSPs) are denoted, provide unprecedented laboratories for studying the cold strongly interacting matter. While the majority of pulsars have a spin period of 0.1 to 1 seconds, MSPs spin with frequencies up to several hundred Hz, thus having a period of milliseconds (ms), giving them their name. The currently accepted formation scenario for MSPs involves an accretion-induced spin-up mechanism: during a long phase of mass transfer in a close binary system of a neutron star (NS) and a companion star, the NS accretes material from its companion, thereby gaining angular momentum and being spin up to an extremely rapid rotation period. As discussed in previous studies ~\cite{1998Natur.394..344W,Guillot:2019vqp}, MSPs are characterized by comparatively weak surface magnetic fields of the order of $10^{8-9}\mathrm{G}$. This property is consistent with their advanced evolutionary stage, since the magnetic field decays over gigayear timescales naturally explains the correlation with their typical age of about $10^{9}$ years. 

The so-called Kepler (mass-shedding) limit establishes the theoretical upper bound on the rotation rate of NSs. It corresponds to the frequency at which the centrifugal acceleration equals to the gravitational pull, such that any further increase in spin would result in the ejection of matter. The current observational record for NS rotation is held by PSR J1748–2446ad, with a spin frequency of 716 Hz~\cite{Hessels:2006ze} followed by the object PSR 1820-30A~\cite{Jaisawal_2024,Guver:2010td} with about the same value. This so far limiting value naturally raises the question of whether the observed frequency approaches the theoretical Keplerian limit. On the other hand, the lack of pulsars spinning significantly faster may instead point out to additional internal constraints, such as the presence of a deconfinement phase transition in the stellar core, which could alter the star’s rotational stability. Therefore, modeling MSPs, including exotic degrees of freedom, as quarks or hyperons, is of great interest. While in Refs.~\cite{Ippolito:2007hn,Dhiman:2010imr, Ayvazyan:2013cva,Bhattacharyya:2017tos,Largani:2021hjo}, hybrid EoSs have been examined in theoretical studies, this field of research recently got new insights and results, as from the Neutron star Interior Composition Explorer (NICER) observations~\cite{Salmi:2024aum,Dittmann:2024mbo,Choudhury:2024xbk,Vinciguerra:2023qxq,Miller:2019cac} and observational data of rapidly rotating pulsars\footnote{The complete pulsar catalog with mass measurements can be accessed at \url{https://www3.mpifr-bonn.mpg.de/staff/pfreire/NS_masses.html}}, motivating us to reinvestigate this topic. In order to explore observable properties of MSPs, an equation of state (EoS) relating thermodynamic properties of cold strongly interacting matter needs to be provided. 

Here, we base our studies on a hybrid EoS, developed in Refs.~\cite{Gartlein:2023vif,Gartlein:2024cbj} and will be discussed in more detail in the next section. The article is focused on the impact of rotation~\cite{1983ApJ...272..702S} on observable quantities such as the oblateness, mass, radius, and angular momentum of NSs. In addition, we employ existing observational constraints to further limit the parameter space of the adopted model. 

\section{Rotating hybrid stars}
\label{sec1}

\noindent Our study is based on the hybrid EoSs presented in Gärtlein et al.~\cite{Gartlein:2023vif}. The low-density region is described by the DD2npY-T EoS, which includes nucleonic and hyperonic degrees of freedom, reproduces the key properties of the nuclear matter ground state, and remains consistent with chiral effective field theory.~\cite{Shahrbaf:2022upc}. The phase transition between quarks and hadrons is modeled by applying the Maxwell construction. The curves of pressure as a function of the chemical potential $p_i({\mu})$ for hadron and quark matter intersect at a certain $\mu_{onset}$. Since the pressure of the quark EoS will dominate after $\mu_{onset}$, the hadronic EoS is replaced by the quark EoS at higher chemical potentials. The latter is based on a confining relativistic density functional approach in the two-flavor color-superconducting (2SC) case~\cite{Ivanytskyi:2022oxv,Ivanytskyi:2022bjc}. The strength of the vector repulsion $\eta_V$ and the diquark pairing $\eta_D$ are the two parameters of the model. Higher values of $\eta_V$ give rise to stronger repulsion between quarks, and, consequently, stiffer EoSs. While an increase of the value of $\eta_D$ leads to an earlier onset of quark matter. Thus, varying the dimensionless $\eta_V$ and $\eta_D$ parameters enables the generation of hybrid EoSs that exhibit a broad range of quark matter onset densities together with a varying stiffness. Since hadronic EoSs that include hyperons and $\Delta$ baryons fail to support neutron stars with masses above 2.1 $M_{\odot}$ due to the significant softening of the EoS~\cite{Tolos:2020aln}, a transition to quark matter provides a physically motivated mechanism to restore compatibility with observational constraints.

The properties of corresponding static spherically symmetric objects in gravitational equilibrium are obtained by solving the Tolman-Oppenheimer-Volkoff (TOV) equations. Besides the boundary conditions, the EoS is required as an input to these equations. The resulting $M-R$ curves are shown in Fig.~\ref{M-R} as solid curves.

To model rapidly rotating NSs, we have adopted the axisymmetric spacetime metric shown below
\begin{eqnarray}
	\label{eqI}
	ds^2=e^{2\nu}dt^2-e^{2\psi}(d\phi-\omega dt)^2-e^{2\lambda}(dr^2+r^2d\theta^2).
\end{eqnarray}
The metric functions $\nu,\psi, \omega$ and $\lambda$  only depend on $r$ and $\theta$. The choice, $e^{\psi} =r^2 \sin^2\theta B^2(r,\theta)e^{-\nu}$ highlights the difference relative to the static NS metric. In this case, equations analogous to the TOV equations should be derived. With the new metric given above, the Einstein equations need to be solved under similar conditions. Although the inputs from the hybrid EoS remain the same as in the static case, the energy–momentum tensor $T^{\alpha \beta}$ through which the hybrid EoS enters—will be modified due to the new metric components. The resulting equations are too complex to be written in closed analytical form. Therefore, the properties of rotating NSs at different frequencies up to the mass-shedding limit are numerically solved via the Rotating Neutron Star (\textit{RNS})\footnote{\url{https://github.com/cgca/rns}} code~\cite{Stergioulas:1998hx}. To explore the different hybrid star properties, we perform the analysis for the two different values of the vector repulsion strength, i.e., $\eta_V=0.30$ and $\eta_V=0.452$, and a set of $\eta_D$ values. Those values are chosen to provide an agreement with the heavy pulsars~\cite{Romani:2021xmb,Romani:2022jhd,Padmanabh:2023vma,Romani:2015gaa} and account for the possibility of the early onsets of quark cores in the NS interior. 

\section{Results}
\label{Res}

\begin{figure}[t]
	\centering
	\includegraphics[scale=0.3]{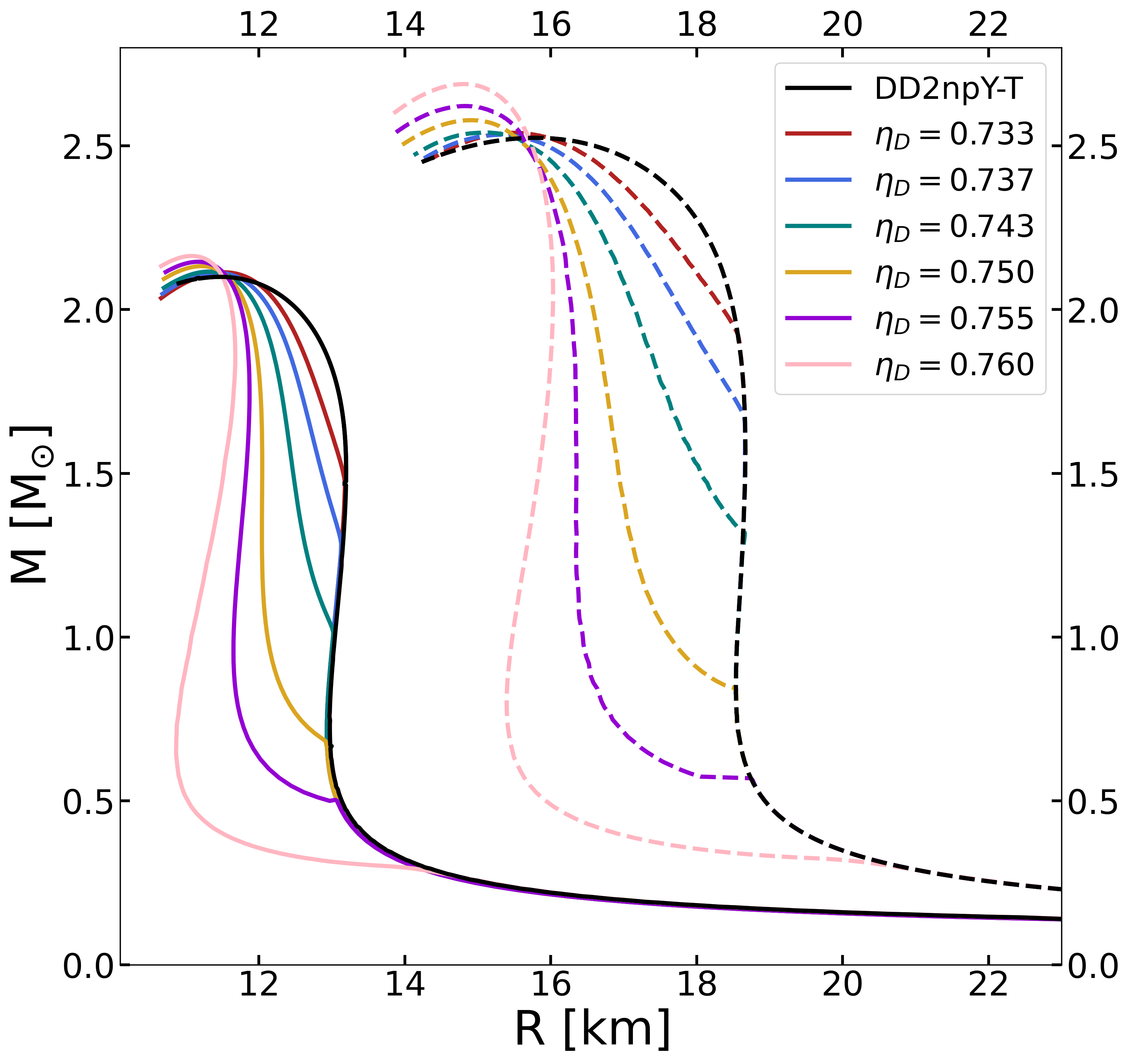}
	\caption{Mass-radius diagram for a set of static (solid curves) and rotating stars with the Kepler frequency (dashed curves). The black and color curves depict the baryonic DD2npY-T EoS, while the color ones show hybrid stars for the fixed vector coupling $\eta_V=0.30$ and different values of the diquark coupling $\eta_D$. The radius corresponds to the equatorial radius. The allowed configurations for hybrid stars are located between the solid and dashed curves of the same color. This figure is adapted from the work of G\"artlein et al. (2025)~\cite{Gartlein:2024cbj}.}       
\label{M-R}
\end{figure}

\begin{figure}[t]
	\centering
	\includegraphics[scale=0.20]{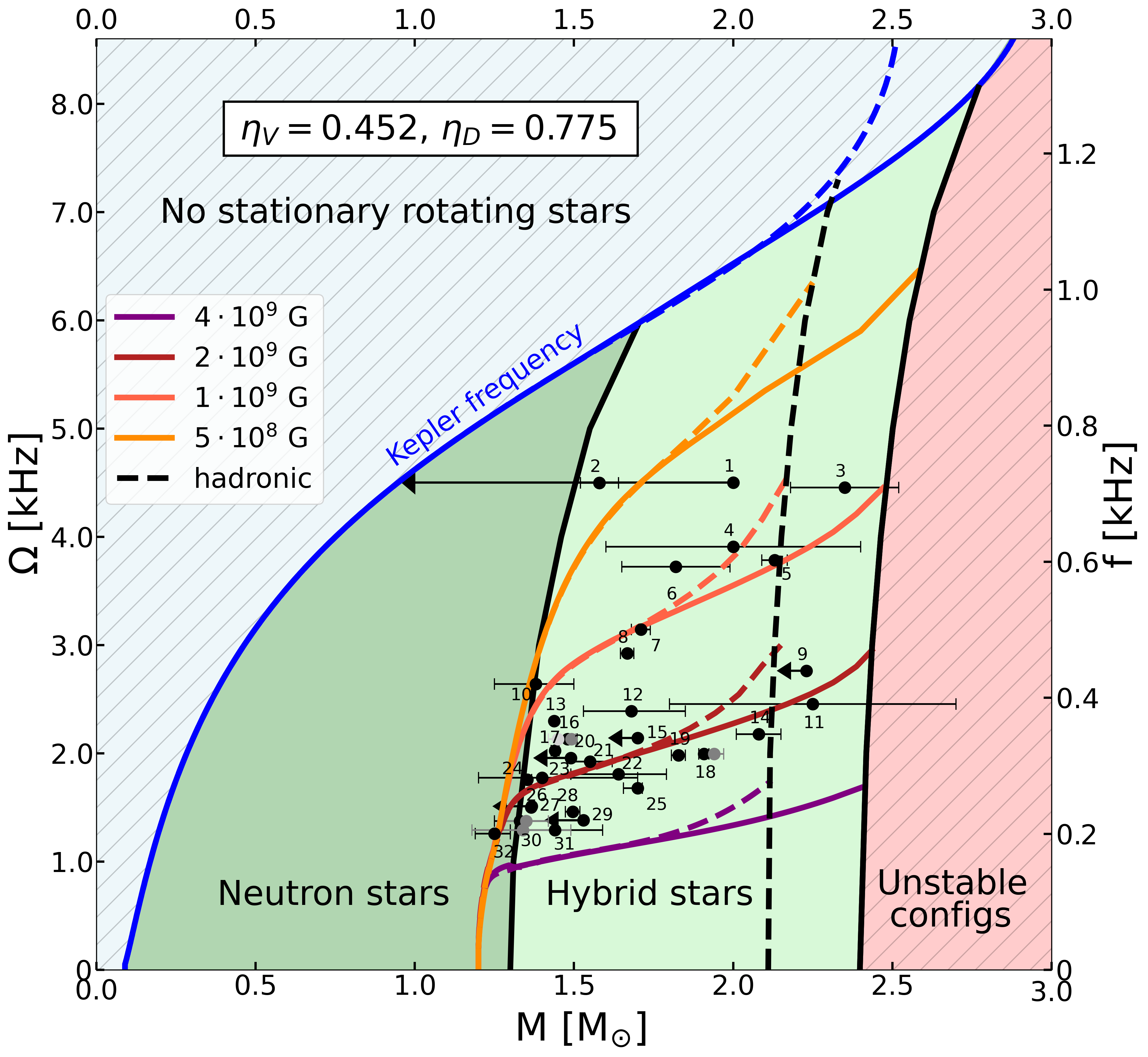}
	\caption{The angular velocity as a function of the gravitational mass of compact stars. The solid blue curve represents the Kepler frequency, above which lies the gray-shaded area where no stationary rotating stars can be found. Black solid curves mark the separation between NSs (dark green area), hybrid stars (light green area), and black holes (pink beige area of unstable configurations). In comparison, the dashed curves represent the purely hadronic analogues. The circles with 1$\sigma$ confidence level error bars depict the measured mass and frequency of the most rapidly rotating MSPs with spin frequency f>200 Hz (the data are listed in Ref.~\cite{Gartlein:2024cbj}). The arrows, rather than error bars, represent the available upper limit on the mass estimate. Other independent mass measurements of the same object are depicted with gray dots. The color solid curves represent the evolution trajectories of the hybrid (solid) and hadronic (dashed) star of 1.2$M_{\odot}$ with the corresponding strength of the magnetic field. This figure is adapted from the work of G\"artlein et al. (2025)~\cite{Gartlein:2024cbj}.}
    \label{om-M}
\end{figure}

\begin{figure}
	\begin{minipage}[t]{0.48\textwidth}
		\includegraphics[width=\textwidth]{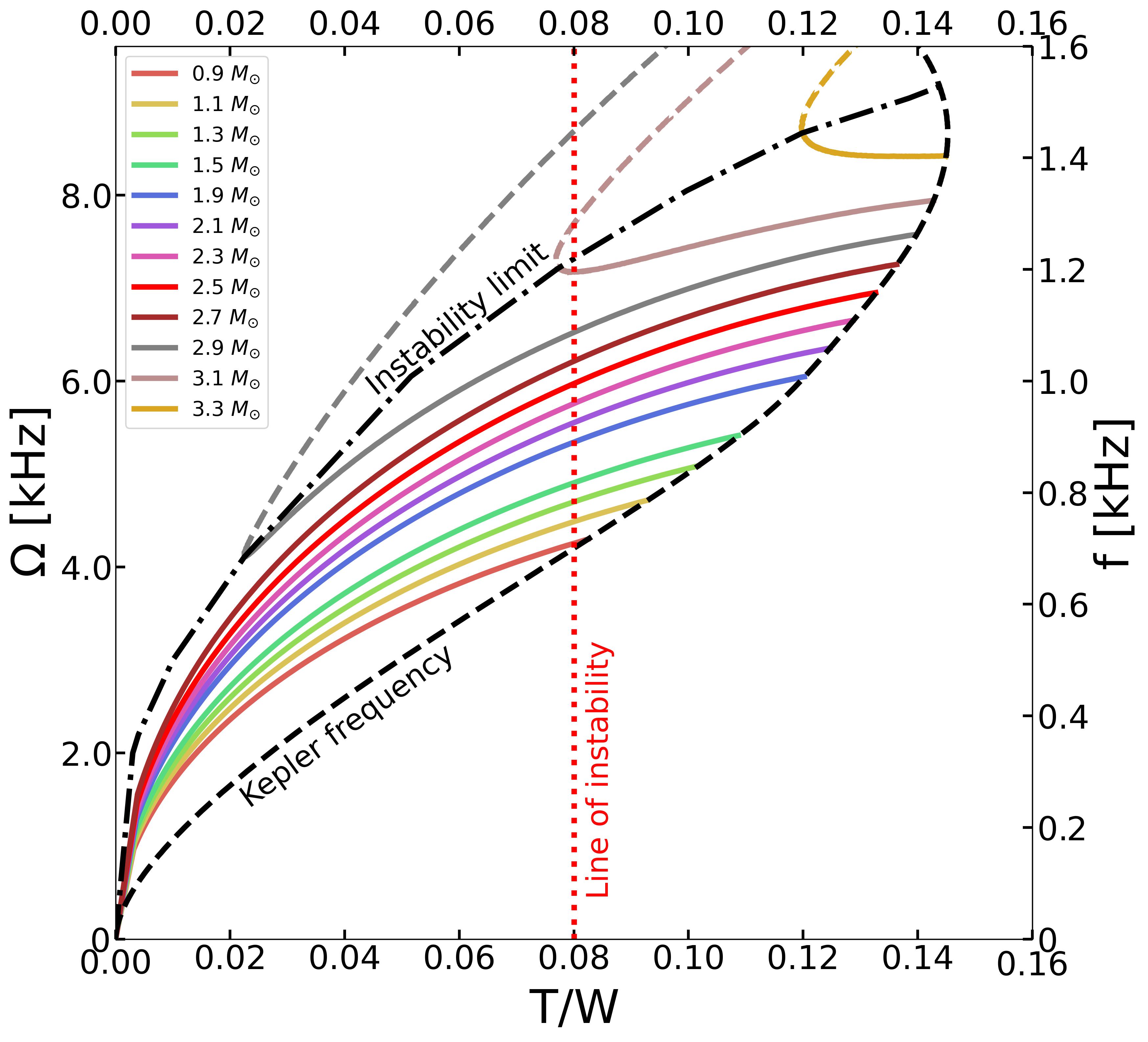}
	\end{minipage}
	\begin{minipage}[t]{0.51\textwidth}
		\includegraphics[width=\textwidth]{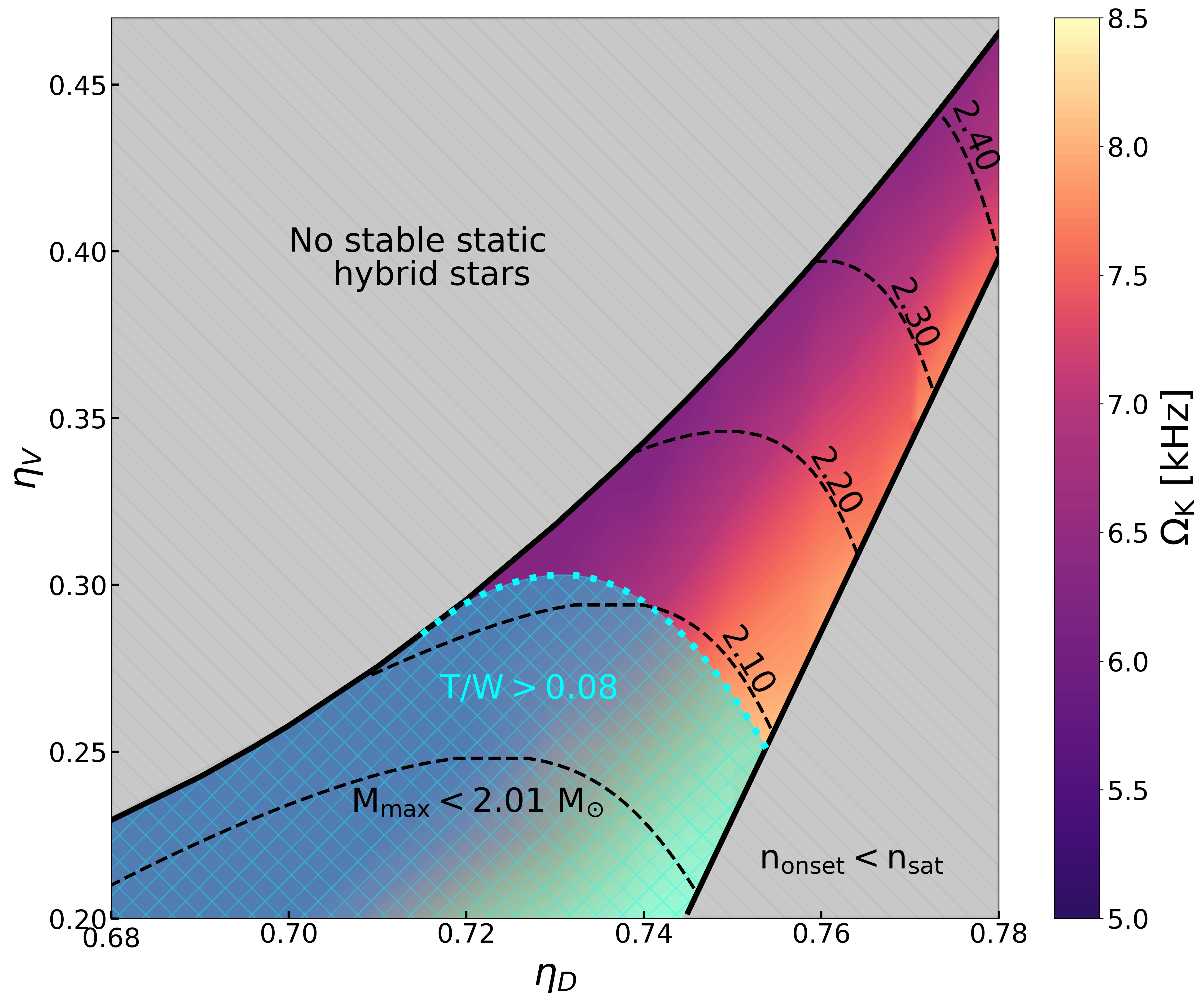}
	\end{minipage}
	\caption{\textbf{Left panel:} Angular velocity $\Omega$ as a function of the ratio of rotational and gravitational energy. \textbf{Right panel:} Parameter plot of the hybrid EoS constrained via the $\Omega-T/W$ plot. This figure is adapted from the work of G\"artlein et al. (2025)~\cite{Gartlein:2024cbj} and shows an additional excluded (blue) area in the $\eta_V-\eta_D$ plain. The physical parameter space is constrained to the colorful area.}
	\label{T-W}
\end{figure}

\noindent 
The EoSs for sets of parameters ($\eta_V=0.30$, $\eta_D=0.733,0.737,0.743,0.75, \\ 0.755,0.76$) were chosen to allow for stars of up to 2.2 $M_{\odot}$ in the static case and a range of different onset densities of quark matter in the NS cores. Fig.~\ref{M-R} shows the static (solid curves) and rotating with the Kepler frequency (dashed curves) $M-R$ relations of the considered set of hybrid (color curves) and hadronic (black curves) EoSs, respectively. 

As seen in Fig.~\ref{M-R}, the general structure of the $M-R$ curves is preserved at the highest possible rotation rate. Clearly, the centrifugal force as a result of the high rotation allows for pulsars with higher maximum mass and much greater radii compared to the static configurations. At the same time, the onset density of the deconfinement phase transition and the special point, at which the set of $M-R$ curves intersect~\cite{Gartlein:2023vif}, are preserved.

We revisit the empirical relation between the Kepler frequency, gravitational mass, and radius of non-rotating NSs, $f_K= C \bigg{(}\frac{M}{M_{\odot}}\bigg{)}^{1/2}\bigg{(}\frac{R}{10\,{\rm km}}\bigg{)}^{-3/2}$. In contrast to earlier studies, which considered only purely baryonic or purely quark compositions, we introduce a new parametrization of the coefficient $C$ that captures different scenarios for the onset of deconfinement while reproducing the two limiting cases of hadronic and quark equations of state. Using the rotation frequency of the fastest known pulsar, PSR J1748–2446ad, at 716 Hz~\cite{Hessels:2006ze}, we derive an updated lower bound on the mass-radius relation of compact stars. This yields an upper limit on the radius of a 
1.4$M_{\odot}$ NS, $R_{1.4} \le 14.90~{\rm km}$, and of a 0.7$M_{\odot}$ star, $R_{0.7}<11.49$~\cite{Gartlein:2023vif}.

In the case of ($\eta_V=0.452$, $\eta_D=0.775$), which is shown in Fig.~\ref{om-M}, we explored the diagrams of angular velocity $\Omega$ as a function of gravitational mass $M$. Interestingly, we can consider scenarios where all included MSPs are hybrid NSs or only the heavier ones include a quark core (see~\cite{Gartlein:2024cbj}). Comparison with the observational data of MSPs (see the table in~\cite{Gartlein:2024cbj}) suggests that the hybrid star scenario provides a more consistent description than purely hadronic models. Among the over 30 MSPs with rotation frequencies above 200 Hz, the Black Widow pulsars~\cite{Romani:2022jhd,Padmanabh:2023vma,Romani:2015gaa} with the highest observed mass (it corresponds to the object 3 in Fig.~\ref{om-M}), almost exclusively show consistency with the allowed region of rotating hybrid stars. 

Since MSPs are thought to form in low-mass X-ray binaries via angular-momentum transfer from a companion star, undergoing a `recycling' spin-up driven by accretion, we employ the accretion model of Ref.~\cite{Poghosyan:2000mr} to study their evolutionary tracks. The color curves in Fig.~\ref{om-M} depict the evolutionary paths of a 1.2$M_{\odot}$ star for the typical accretion rate of $10^{-7}M_{\odot}/{\rm yr}$, a decay time $\tau_B=10^8$ yr of the magnetic field and different values of the magnetic field amplitude shown in the legend. The dashed and solid curves represent the purely hadronic and hybrid configurations, respectively. 

As seen in Fig.~\ref{om-M}, the obtained curves provide a good agreement with the intriguing clustering of MSPs around $\sim 2\mathrm{kHz}$. According to our model, it occurs in the vicinity of the phase transition line, which can be interpreted as a consequence of the interplay between accretion-driven spin-up and an internal phase transition. As pointed out in Ref.~\cite{Blaschke:2008na}, the formation of a quark core can give rise to a waiting-time phenomenon, potentially explaining the increased number of MSPs in this region. \\

\noindent 
We briefly discuss our analysis of the stellar oblateness $e=\sqrt{1-(R_p/R_{eq})^2}$, which quantifies the degree of flattening of the star due to rotation, i.e., the relative reduction of the polar radius $R_p$ compared to the equatorial radius $R_{eq}$. The stellar spin evolution may cause a phase transition in the stellar interior, transforming a hybrid star into a standard NS or vice versa. We therefore investigated the behavior of the oblateness as a function of the angular velocity $\Omega$, assuming a transition from a hybrid to a purely hadronic configuration. As shown in our original work~\cite{Gartlein:2024cbj}, the oblateness remains a smooth and continuous function of $\Omega$ even across the composition change. 

\noindent 
On the left panel of Fig.~\ref{T-W}, we show the angular velocity $\Omega$ as a function of the ratio of rotational to gravitational energy $T/W$ for the same pair of coupling constants as in Fig.~\ref{om-M}. The results demonstrate that for high values of rest mass, no static NS solutions exist. These only survive under sufficient rotation and collapse if their rotational velocity decreases. Most importantly, the vertical red dotted line depicts the onset of non-axisymmetric f-mode instabilities~\cite{1985ApJ...294..463M,1986ApJ...304..115F,Morsink:1998db}. Above the value of $T/W\approx 0.08$, the non-radial instabilities lead to gravitational wave radiation. Thus, a higher ratio will cause the rotating NS to collapse, ruling out the existence of any rapidly rotating massive pulsar configurations with ratios exceeding 0.08. Applying this constraint to the whole parameter space of the present model, we can exclude the area shown in blue on the right panel of Fig.~\ref{T-W}. The colorful area was already presented in Ref.~\cite{Gartlein:2023vif} as allowed configuration space. When the constraint derived from $T/W$ is included, the physically relevant set of $(\eta_V,\eta_D)$ pairs becomes even more restricted.

\section{Conclusions}
\label{conc}

\noindent Rotating NSs, and in particular the MSPs observed to date, continue to play a key role in constraining the properties of dense matter. We demonstrated that the most rapidly spinning and massive Galactic NS, PSR J0952–0607, can be explained as a hybrid star with a color-superconducting quark matter, while such properties remain unattainable for a purely hadronic model with hyperons. We argue that a transition to quark matter offers a physically well-motivated mechanism to reconcile the EoS with observations of heavy NSs, overcoming the substantial softening caused by the appearance of hyperons and $\Delta$ baryons.

We revise the empirical Kepler frequency relation by introducing a new parametrization of the factor $C$ that accounts for different possible deconfinement scenarios while reproducing the limiting hadronic and quark cases. Using the 716 Hz spin of PSR J1748–2446ad, we obtain updated constraints on the mass–radius relation, yielding upper limits of $R_{1.4} \le 14.90~{\rm km}$, and $R_{0.7}<11.49$ km.

In addition, it has been shown that introducing a simple model of accretion and magnetic-field decay provides a realistic explanation for the observed clustering of pulsars around 2 kHz and for their evolutionary paths. This clustering occurs near the phase transition, acting as a saddle point in the pulsars’ evolutionary trajectories, and may be a consequence of the onset of quark cores within compact stars.

Moreover, we investigated the angular momentum, oblateness, and the ratio of rotational and gravitational energy as a function of the angular velocity (for a more extended analysis, see Ref.~\cite{Gartlein:2024cbj}). Notably, the oblateness remains smooth across the compositional transition. In addition, the study of $T/W$ allowed us to further constrain the physical space of model parameters applied.

In summary, this work provides a further significant contribution to the investigation of MSPs as potential hybrid NSs and substantially complements previous studies in this area. With the advancement of future observational surveys and programs, such as the Square Kilometer Array Observatory (SKA)~\cite{Watts:2014tja}, the increased precision and number of observed pulsars might give rise to further insights into the internal composition of NSs. 

\section{Acknowledgments}

C.G. and I.L. express their gratitude to the Funda\c c\~ao para a Ci\^encia e Tecnologia (FCT), Portugal, for providing financial support to the Center for Astrophysics and Gravitation (CENTRA/IST/ULisboa) through Grant Project No. UIDB/00099/2025. C.G. also acknowledges the Funda\c c\~ao para a Ci\^encia e Tecnologia (FCT), Portugal, through the IDPASC PT-CERN program with the No. PRT/BD/154664/2022. This work of O.I. was performed within the program Excellence Initiative--Research University of the University of Wrocław of the Ministry of Education and Science and received funding from the Polish National Science Center under grant No. 2021/43/P/ST2/03319. V.S. gratefully acknowledges support from the UKRI-funded ``The next-generation gravitational-wave observatory network'' project (Grant No. ST/Y004248/1). This work was produced with the support of INCD and funded by FCT I.P. under the Advanced Computing Project \\
2023.10526.CPCA.A2 with DOI identifier 10.54499/2023.10526.CPCA.A2.
D.B. acknowledges the support from the Polish National Science Center under grant No. 2021/43/P/ST2/03319. 

\bibliographystyle{elsarticle-num}  
\bibliography{references}

\end{document}